# Wireless Communication in Biosystems

Jingjing Xu, Fan Yang, Danhong Han and Shengyong Xu*

Key Laboratory for the Physics & Chemistry of Nanodevices, and Department of Electronics, Peking University, Beijing, 100871, P. R. China

* Correspondence author: xusy@pku.edu.cn

**Abstract:**

Electrical signals play the primary role in rapid communication among organs, tissues and cells in biosystems. We have analyzed and discussed three phenomena of electrical communication in biosystems, including direction-changing movement of paramecia, prey behavior of flytraps, and simultaneous discharge of electric eels. These discussions lead to a conclusion that in biosystems the electrical signals are mainly soliton-like electromagnetic pulses, which are generated by transient transmembrane ion currents through protein ion-channels. These electromagnetic signals mainly propagate along the softmaterial waveguide which is basically composed of dielectric membrane, sandwiched in two ionic liquid layers at both sides. In myelinated axons, the dielectric layer is as thick as several microns as it consists of tens of lipid bilayers. This transmission model implies that a wireless communication mechanism has been naturally developed in biosystem. This hypothesis may shed some light on the working mechanism of ten billion neurons in a human brain.

## 1. Introduction

For lives developed from nonorganic materials, in addition to many important factors such as a suitable environmental temperature range, input of external energy (sunshine as on earth) and establishment of proper metabolism processes, there is a crucial requirement. To make a life system dynamically stable and sustainable, it needs to involve a huge amount of atoms and molecules working in synchronization. For instance, a single cell consists of a few hundred trillion atoms, and an organ is often built up with millions and billions of cells. The increasing number of involving atoms, molecules, cell organelles and cells may greatly enhance the capability of a life system to accomplish complex functionalities and behaviors, however, as a tradeoff it also greatly increases the difficulty in communication among the vast unites, as well as in organization of the whole system.

In general, a life system utilizes both electrical and chemical carriers for inner communication. Molecules carrying chemical information, such as hormones, move slowly in liquid environment of a life system. Therefore, for fast activities such as the motion of a body, electrical signals are usually the only choices for message carriers. Several theories have been developed to understand the electrical communication in a neural system. The Hodgkin and Huxley model developed from experiment data of giant squid axons, has been well accepted for understanding the generation and transmission of electrical signals in nerve system [1-4]. This theory describes how a pulsed electric signal, often referred as "action potential", a transmembrane voltage pulse of approximately 100 mV propagates along the cable - neuronal axon. This measurable voltage pulse results from a transmembrane flows of $Na^+$ and $K^+$ ions through protein channels when they are triggered open [2]. The theory describe the transmission behavior of the action potentials with a circuit model based on equivalent resistors (ion channels and the cytosol of the axon) and capacitor (lipid membrane). However, this model faces difficulty in explaining some phenomena such as "*saltatory propagation*" in myelinated axons, or cross-over of two action potentials on one axon which appears missing the "*refractory period*".

One decade ago, a novel electromechanical soliton model was presented [5]. Based on evidences of reversible heat changes, thickness and phase changes of the membrane observed during the action potential [6-12], Heimberg and Jackson describes the transmission of electrical signals in nerve systems in the form of mechanical dilatational waves, which are generated by transmembrane ion low induced heat. The waves propagate like electromechanical solitons and depend much on the presence of cooperative phase transitions in the membrane. The model is applied to explain most known features of nerves, including problem of refractory period, reversible release and reabsorption of heat, effect of anesthetics [13], as well as emergence of ion channel phenomena from thermodynamics of the membrane without employing molecular features of the membrane components [14-18]. The model also carried out simulations showing that the minimum velocity of the solitons is close to the propagation velocity in unilamellar vesicles [5]. Recently, Xue and Xu stated that indeed it is soliton-like electromagnetic (EM) pulses rather than a mechanical wave that transmitting within the lipid membrane. The thin lipid bilayer membrane of an unmyelinated axon, or the thick sheath of a myelinated axon, together with electrolyte solutions on both sides, form a kind of "softmaterial waveguide" for efficient transmission of the EM pulses [19]. This hypothesis can be applied to explain various electrical communication phenomena in nerve system, as well as electrical communication phenomena in many other circumstances without nerve systems [19, 20].

In addition to pulsed electric signals, indeed various weak photon emissions so called "biophotons", have been detected since 1920's in bacteria, plants, animal cells, even in the nerve systems of human beings [21-23], and it has been suggested that biophotons play an important role in electrical communication of biosystems [24, 25]. Thar *et. al.* believed that this kind of biophotons could be guided along a filamentous mitochondrial network, where the microtubules acted as optical waveguides inside neurons. They gave the model and theoretical calculation, where the core of microtubules were taken as the transmission path for biophotons [26].

Yet to date some important questions remain open. For instance, how does the electrical signals exchanges among a large number of cell organelles or among cells? How does electric communication work out in life systems without neurons, such as in a plant? And, how can billions of ion channels open simultaneously within 0.1-0.5 ms in an electric eel [27, 28]? To answer these questions, one may seek help from quantum entanglement at room-temperature, or even unknown physical mechanisms. We show in the paper that these problems can be solved with conventional physics. By analyzing several systems where a large amount of effectors (e.g., proteins or cells) respond synchronously to a single source, including paramecium, flytrap and electric eel, we reveal a universal electrical communication model existing in biosystems with or without neurons and axons. In this communication model, soliton-like electromagnetic pulses serve as the message carriers, and membranes based softmaterial waveguide networks serve as the transmission paths.

## 2. Electrical communication in different biosystems

*2.1 Electrical communication in a paramecium*

We first discuss electrical communication phenomena in unicellular biosystems. Paramecium is a representative unicellular organism. Unicellular organisms are thought to be the oldest form of life, with early protocells possibly emerging 3.8–4 billion years ago [29]. A paramecium, although being a single cell, has function units for swimming, for food in-taking, digesting, and for reproduction [30-32]. A paramecium, with a size of 30-200 μm in total length, has as many as 5000-6000 cilia for its swimming movement. Each cilia is about 10 μm long and 250 nm wide [33]. A paramecium swims in water by synchronously beating its cilia forward or backward. It is an interesting phenomenon, that when a paramecium touches a subject on its way swimming forward, it is able to reverse its swimming direction very quickly by changing the beating direction of all its cilia simultaneously [33, 34].

Roger Eckert *et al.* found in their experiments that it just needed 30 ms for all 5000-6200 cilia of a 200 μm long paramecium to reorient their beating direction [35]. This leads to a propagation speed of the signal about 6.8 mm/s from one end where an external stimuli is applied to the far end of the paramecium. As all the cilia are embedded on the same cell membrane, it is not likely that chemical signals carried by molecules play the role for realization of synchronized motion of the cilia over the whole paramecium. Not only because molecules diffuse slowly in liquid, but also it needs identical molecules reaching each cilia within a short time. Mechanical motion of the membrane, as presented by Thomas Heimberg and Andrew D. Jackson in explaining propagation of action potentials in axons [13-18], may be applied to understand the synchronized reaction of thousands of cilia in changing their beating directions. However, we believe that a paramecium has used electrical pulses to fulfill this mission in processes discussed as the following.

Previous studies have shown that, when the impact occurs at the front side of a paramecium, it increases the permeability for $Ca^{2+}$ and leads to an inward flow of $Ca^{2+}$ at anterior membrane, thus resulting reversal motion of all the cilia. This initiates the "avoidance response" [33, 34, 36, 37]. While an impact at the caudal (posterior) surface of a paramecium, it evokes an increasing permeability of $K^+$ and causes an outward flow of $K^+$, thus enhancing the frequency of beating of all the cilia, which makes the paramecium swim faster to escape [33, 34, 36, 37].

In our model, when a paramecium runs into an obstacle on its way, the receptors near the impact point sense the stimuli and generate electromagnetic pulses. The function of the receptors is similar to those mechanical sensors (e.g., tactile corpuscles) in advanced animals [38-41], where the sensor converts deformation in its proteins caused by the mechanical impact into transmembrane ion flows of $Ca^{2+}$ and/or $K^+$. As discussed previously, these transient transmembrane ion flows introduce directly soliton-like EM pulses [19, 20]. Here, the membrane system of a paramecium, together with ionic solutions inside and outside of the membrane, serves as the effective transmission path for the EM pulses. Once EM pulses are generated by the receptors near the impact point, the EM pulses would propagate quickly along the membrane system all over the paramecium, and it is the change of local electrical field near each cilia that triggers the change of its motion mode. The EM pulses travel in a dielectric medium at a speed more than one tenth of light speed [19, 20]. The time for changing swimming direction of a paramecium, which is 30 ms observed in some experiments [35], mostly costs on the change of mechanical beating mode for the cilia. As a result, thousands of cilia reorient almost simultaneously their beating direction and frequency.

These processes are schematically illustrated in Figure 1.

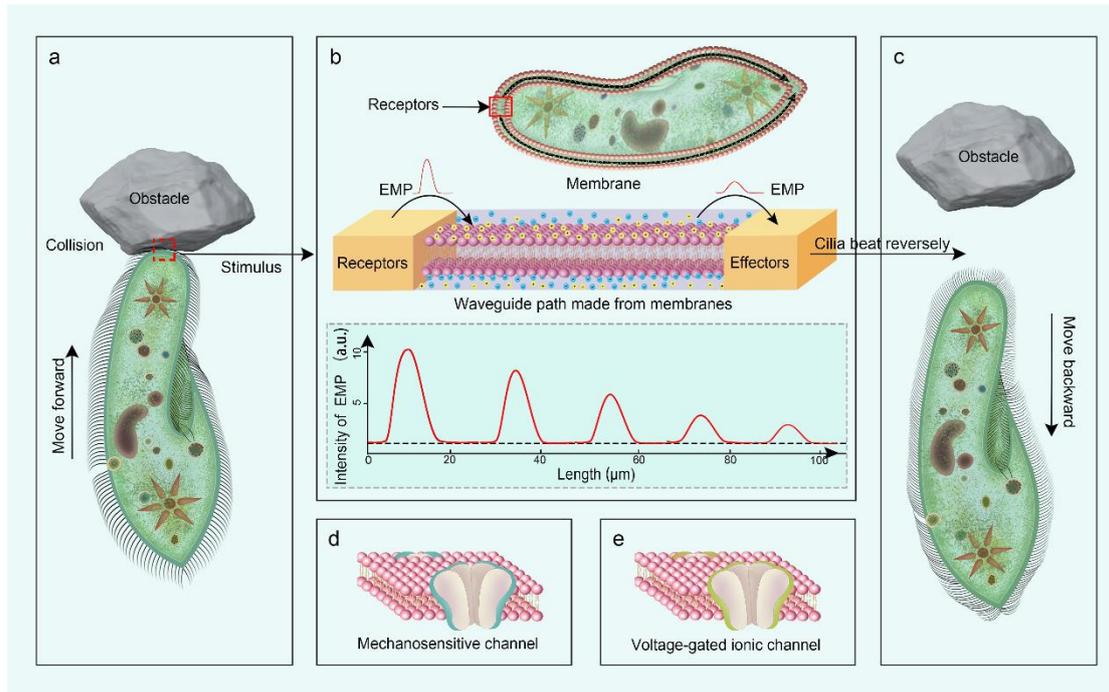

Figure 1. The electrical communication approach in *Paramecium*. (a). The anterior of a paramecium is collided by an obstacle when moving forward. (b). The receptors near the impact point sense the stimuli and generate EM pulses, which transmit throughout the whole body via the membrane-based waveguide, and trigger all effectors simultaneously. During transmission, the intensity of EM pulses (EMP) decreases attenuates over distance. (c). Effectors receive the EM pulses and make the associated cilia change their beating mode, leading to reverse motion of the paramecium thus moving away from the obstacle. (d). Possible receptors on somatic membrane – mechanosensitive channels, which could generate EM pulses by transmembrane transient ionic current. (e). Conceivable effectors on ciliary membrane – voltage-gated ionic channels, which are triggered open with the steep change in electric field when the EM pulses passing through, thus they induce secondary ion currents and change the relative movement mode between microtubulins in cilia.

It has also been observed that there are many mechanosensitive channels on the somatic membrane of a paramecium, and voltage-gated channels on the ciliary membrane [35]. As a result, it is believed that the receptors for sensing impact are the mechanosensitive channels through which the transmembrane transient ionic current bring about EM pulses. And the conceivable effectors are the voltage-gated channels, which can be triggered open by a steep change in electric field when the EM pulses pass through [20], leading to a leak flow of $Ca^{2+}$ ions which change the relative movement between microtubulins in cilia. In view of the length of a paramecium, the EM pulses generated at the impact point are supposed to propagate all over the body through its membrane based waveguide, similar to the cases in unmyelinated axons, where EM pulses generated by ion channels at one spot can trigger the opening of ion channels at neighbor spot located tens of microns away.

One sees from this case that, electrical communication dose exist in unicellular biosystems. Even at the single-cell level, soliton-like EM pulses could play as message carriers and deliver the information along membrane based waveguide structure all over the whole body of a paramecium, resulting synchronized reaction of thousands of effectors (here being cilia in a paramecium).

*2.2 Electrical communication in a flytrap (Dionaea muscipula)*

Next we take a look at electrical communication in more advanced biosystems. Plants are multicellular organisms, but they have not developed neural systems. However, electricity phenomenon in plants like *Mimosa pudica* and *Dionaea muscipula* were observed long ago [42, 43]. For example, in 1873, Burdon Sanderson captured the distinct and precise electric pulse signals on the leaf surface of *Dionaea muscipula* [44]. It was reported that the transmission speed of electric signals in plant span a lot, while the highest speed was surprisingly comparable to that measured in myelinated axons [45], indicating a similar transmission mechanism.

Trapping action of a flytrap is performed by two leaves jointed at the leaf stalk position. Three equivalent sensory hairs with a length of ~ 2 mm and a width of ~ 200 μm stand on the inner side of each leaf. Once any two of the six sensory hairs are triggered twice by a fly or a bug and the time interval is less than 20-30 s, the two leaves would close in 100-300 ms by the reaction of leaf stalk, completing a trapping performance [44, 46-49]. Similarly, in some cases if the six sensory hairs are triggered three or more times within 50 s, "memory effect" is observed and the two leaves could also close and catch a fly [49]. The prey behavior of a flytrap need a fast reaction, where the signal speed from the receptors (under the sensory hairs) to the effectors (located in the leaf stalk) is in the order of 0.2 m/s or faster [45], comparable to that for action potentials propagating in unmyelinated axons.

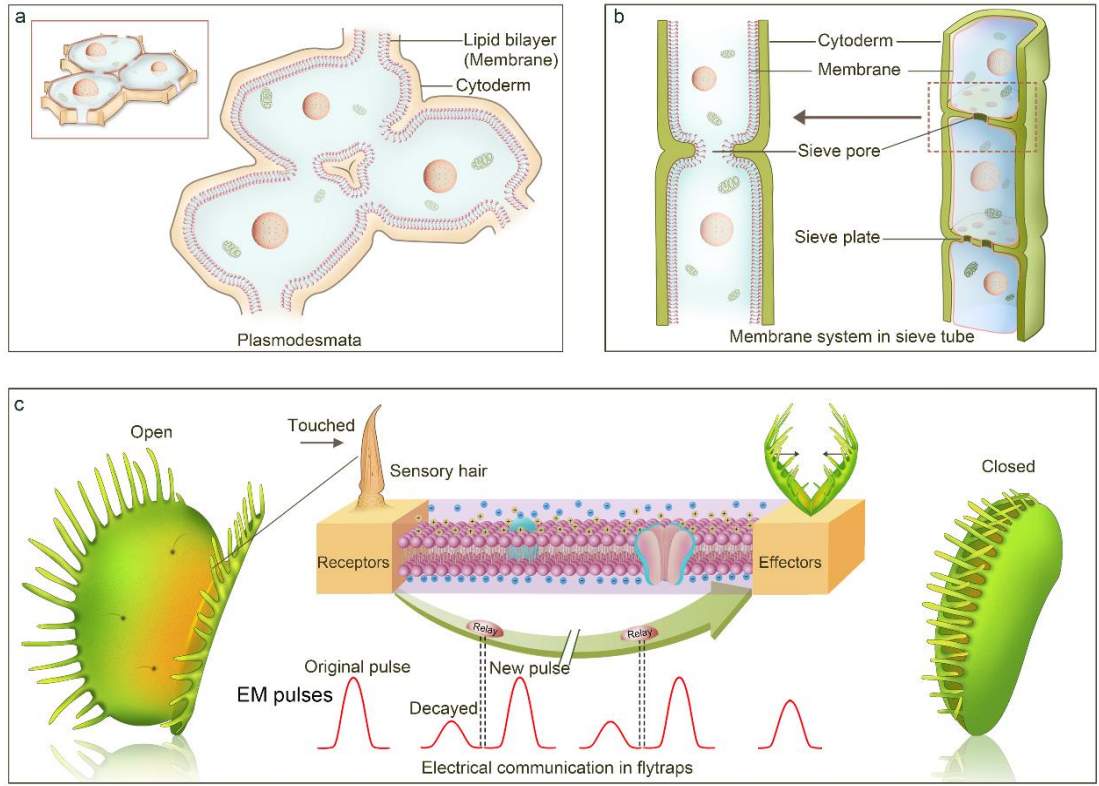

Figure 2. (a). Illustration of an omni-directional waveguide network in a *Dionaea muscipula* built up with cell membranes and walls. (b). Directional membrane network built with sieve tubes. (c). Trapping action and its working mechanism of a *Dionaea muscipula*. The receptors under sensory hairs transfer mechanical shift of the hair into EM pulses via transient transmembrane ion currents, then the EM pulses propagate along the waveguide network embedded in the leaves, with attenuation and several relays on the path to the effectors in the stalk, and finally cause reaction of the stalk and close of the two leaves.

We suggest that, similarly to the case in a paramecium the electrical signals occurred in plants are also in the form of EM pulses. In trapping action of a flytrap, the EM pulses are generated by the sensory receptors for mechanical shift at bottoms of the six equivalent hairs when triggered twice within 20-30 s [50-53]. Evidences have been observed showing that the sensory processes are strongly linked to transient transmembrane inflows of $Ca^{2+}$ [54]. This kind of transient transmembrane $Ca^{2+}$ flow is probably the basic source for EM pulses in plants [19, 20]. We suggest that the propagation of the EM pulses in plants relies on a membrane based waveguide network as schematically illustrated in Figure 2. The main framework of this network is built up with cell membranes and walls of plasmodesma. Plasmodesmata are microscopic channels that connect extended membranes of two adjacent plant cells [45]. In addition, plasmodesma across sieve pore can connect the membrane of adjacent two sieve tube cells, therefore building up a continuous waveguide network along the sieve tube, as shown in Figure 2B.

Previous studies have recognized two forms for electric signal communication in plants [54-56]. Traps of flytraps and some lower plants possess a form of omni-directional propagation, which is similar to that of cardiac myocytes. More common electrical signals in higher plants are directionally propagated in vascular bundles along the plant axis. These two forms are probably corresponding to the two types of membrane network mentioned above respectively.

We suggest that, due to the attenuation in intensity of the EM pulses transmitting in the softmaterial waveguide network and the long distance between receptors and effectors up to one centimeter or more, it needs ion channels serving as "relays" on the transmission path. These relays are triggered by the original EM pulses passing though and generate secondary EM pulses, thus renew the electrical signals. Voltage-gated ionic channels observed in plants may play the role of relays, who can be triggered open by external EM pulses and evoke new EM pulses. This process is very much similar to the transmission processes of action potentials in axons. Here in plants, the cell membrane and wall based network serves as the waveguide path for EM pulses.

*2.3 Electrical communication in synchronous discharge of an electric eel*

Animals, as more advanced biosystems than plants, have develop neural system in the long history of evolution. In myelinated axons, electrical signals in the form of action potentials may propagate at a speed up to 100-150 m/s [57]. In extreme cases such as the discharge activities of electric eels (e.g., *Electrophorus Electricus*), the transmission speed of electrical signals could be even faster. Many studies have focused on the discharge behavior of electric eels when they hunt and cruise in the view of ethology [27, 58-60], and on the role of a single discharge cell (also known as electrocyte) from the perspective of physiology [61-63]. Kenneth Catania found an electric eel could discharge about 400 times per second, showing a period of 2.5 ms and a full-width at half maximum (HWHM) of ~ 0.5 ms for a single peak [27], as shown in Figure 3. In terms of time duration, these discharging peaks are very similar to action potentials in neural systems [28]. Caputi *et al.* showed that all ionic channels participating in discharge in the whole electric organ of the eel opened in 0.1-0.5 ms [64]. A grown eel has an electric organ more than 1 meter, which consists of many layers of electroplate, which is made of thousands of electrocytes at the same level and perpendicular to the longitudinal axis of the fish, each with thickness of 100 microns [63, 65, 66]. As the output peak voltage of a single electrocyte is measured to about 150 mV, an eel needs to cascade 2000 ~ 5000 layers of electroplates to realize a total external voltage of 300-800 V as measured in experiments [58, 65-67]. Meanwhile, an eel can release a total external current of about 1 amp [58], it needs to let $10^{6-8}$ ion channels located on each layer of electroplate in parallel configuration simultaneously discharge. When both factors for cascading in voltage and parallel in current are considered, an eel needs to simultaneously discharge $10^{9-11}$ ion channels within 0.1-0.5 ms to attempt one killing performance [27, 28]. Considering the size and geometric structures of the electric organ, this is an extremely difficult task. For example, if an electrical signal is transmitted from one end to the other end of a 1 meter long electric organ at a speed the same as that for action potentials in myelinated axons, i.e., 100 m/s, it needs 10 ms. This time cost is almost two orders of magnitude higher than the discharge time measured in a real electric eel.

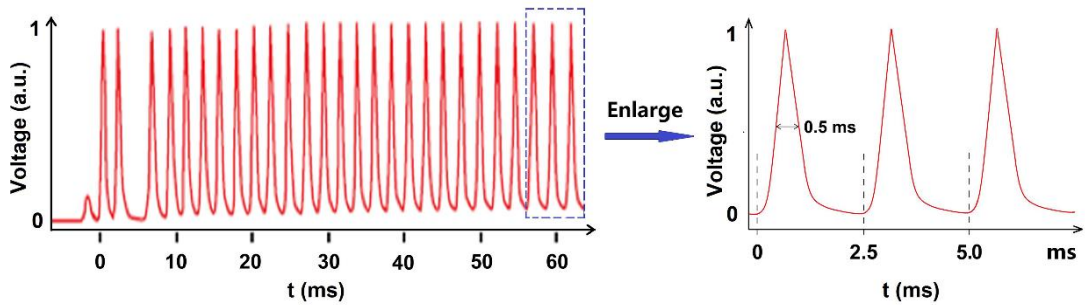

Figure 3 Discharge of an electric eel. The left part modified from Figure 1 in reference [27]showed the frequency of discharge. The right part is the enlarged discharge, indicating that the full-width at half maximum of a single discharge is ~ 0.5 ms.

To data it is still not clear how an electric eel could simultaneously discharge billions of ion channels within less than one millisecond [68-72]. In previous studies, the electromotor system for an electric eel to complete a hunting discharge includes the brain, medullary electromotor nucleus (EMN), large bulbospinal tract (BST) and electrocytes [68, 69, 73-77]. The brain is the command center giving orders for discharging. EMN consists of two types of large neurons (approximately 70 μm and 30 μm in diameter respectively) and it occupies the upper middle region of the spinal cord [78]. EMN is considered as the key for transferring the order to the rostral and caudal simultaneously [68, 69, 73-77]. BST, a wedge of white matter in the dorsal-lateral region of the cord, consists of descending electromotor fibers which convert the orders from the EMN to the electrocytes nearby. A large number of endings of descending fibers act on the posterior half of the electrocytes, with a density of ~ 2000 /mm$^2$ [74], through chemical synapses to open the sodium ion channels. Many researchers have attempted to explain the mechanism for the discharge process of the electric organ. Some believed that the close appositions (electric synapses and surrounding inter-mediate junction) among neurons in EMN might contribute to the synchronization activity of the organ [68, 78]. Others found low velocity fibers (~10 m/s) predominated in the rostral cord while fast velocity fibers (~100 m/s) at distal levels, which could compensate the different conduction distances from the command center [69].

We agree with previous researchers that the EMN in spinal cord should play a key role in synchronization of the discharging process. We propose here a model based on softmaterial waveguide network to explain the working mechanism of the extremely fast discharging activity, as illustrated in Figure 4.

Briefly, when an eel fish makes a decision to take an action, its brain sends a command, such as a series of coded spikes to the synchronizer in EMN through myelinated axons, then synchronizer in EMN sends synchronized orders via descending fibers to thousands of electroplates in the electric organ to perform a short, simultaneous discharge. An electric eel needs to complete two steps of synchronization for a simultaneous firing output of its billions of ion channels. First, it needs to make sure that thousands of electrocytes on the same electroplate work simultaneously. Second, it needs to ensure a simultaneous reaction among thousands of cascaded electroplates from the front to the end of an electric organ.

To realize the first synchronization, the key structure for the first one is a pure membrane based waveguide network in the EMN for transmission of EM pulses. It is well known that EM pulses could travel in dielectric materials at a speed as high as $c/\sqrt{\varepsilon}$, where $c$ is the light speed and $\varepsilon$ is the dielectric constant, 3-4 for the lipid membrane. Therefore for EM pulses transmitting through a 1 meter long spinal cord it takes only $10^{-8}$ s. Unfortunately, EM pulses generated at the first group of ion channels of the synchronizer are supposed to attenuate along the transmission path and it needs relays on the way to recover signals the same as or similar to the original ones. Then the unclear part becomes whether the electric eels have developed unique waveguide structures in EMN where spacing between neighboring relays is in the order of 10 cm, thus it needs only ten recovering processes on the way and costs less than 1 ms for propagation of original signals to the far end of EMN. Indeed, this is corresponding to a transmission speed of 1000 m/s or more, much faster than those measured in normal myelinated axons. In our opinion, the coupling of membrane waveguides (myelin sheath from two adjacent axons) could improve the transmission speed by decreasing the necessary firing times (i.e., by improving the transmission efficiency of EM waves) in the process [19]. Furthermore, the transmission speed difference of 10 times in BST between the rostral and the caudal might decrease the time relay slightly between two ends to 0.1-0.5 ms [69], which is corresponding to the morphology features. In the caudal regions, both the axon diameter and myelin sheath are thicker, and the internode lengths are longer [68, 78]. This favors realization of the synchronization process.

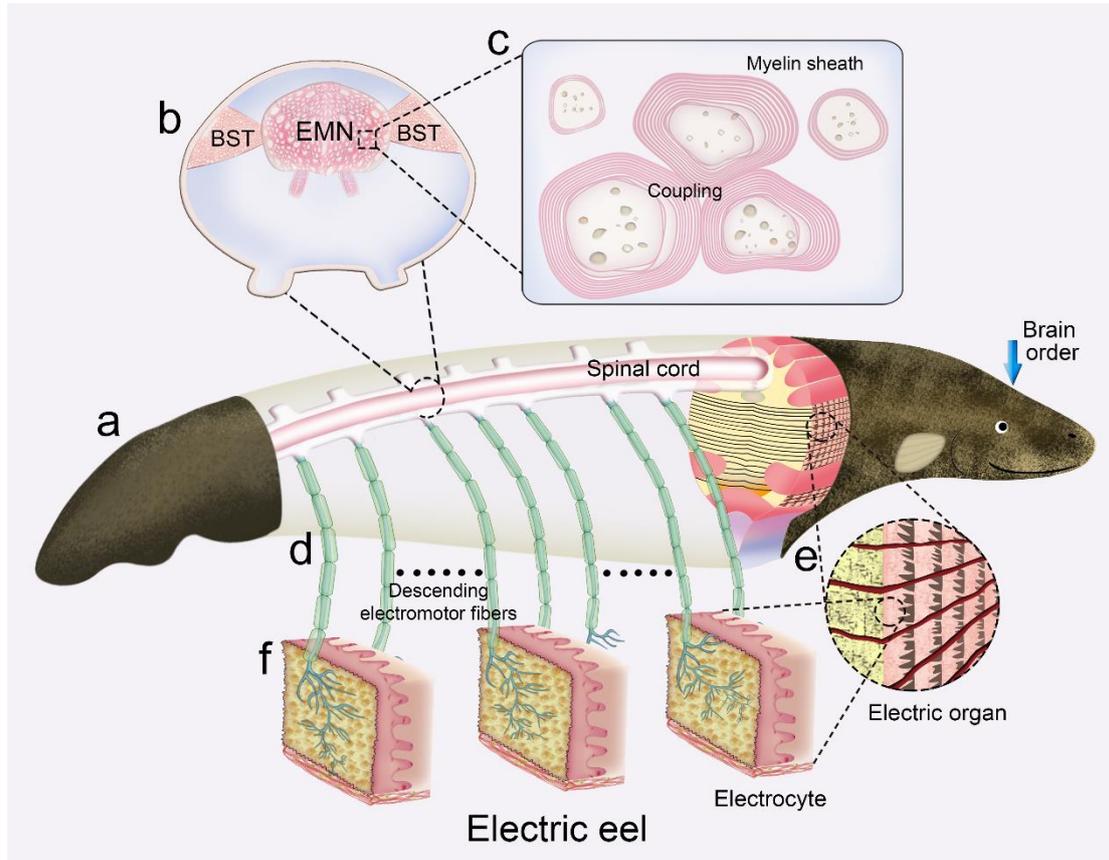

Figure 4. The electrical communication system in first model for synchronized discharge in an electrical eel. (a) The system is made from four parts: the brain, medullary electromotor nucleus (EMN), large bulbospinal tract (BST) including descending fibers, and electrocytes. (b) EMN ensures the signals transmit to the caudal of the spinal within 1ms, playing the main role in synchronization in discharge. (c) The coupling between and/or among the large neurons in EMN based electric synapses and surrounding inter-mediate junctions make sure high efficiency of EM waves. (d) The different features inluding diameter, myelin sheath thickness and internode length between the rostal and the caudal implies the higher speed for descending fibres with longer distance to the rostral, which can slighty compensate the delay. (e) The enlarged picture shows the structure of electric organ. (f) The fibres branch a lot of nerve endings connecting with the posterior membrane of every electrocyte with an approximate density of $2000/mm^2$.

The second synchronization need a way to send out orders simultaneously to all electroplates via these parallel fibers. For the thousands of electrocytes in the same electroplate, the discharge synchronization could be realized by varying the axon diameter, myelin sheath thickness and internode length of parallel myelinated axons. According to previous work, larger diameter, thicker myelin sheath and longer internode length all result in a faster propagation speed of the transmitted electric signals [19]. As all the membrane and myelin sheath of axons are mainly insulating dielectric materials, indeed such a special axon system form a waveguide network for specially developed for the synchronization purpose. As a result, electrocytes with different distance to EMN in spinal could receive the discharging orders almost simultaneously. It is surely a difficult task, but we have seen many extreme cases in nature that the long time evolution could turn many difficult tasks into reality. Indeed, the total delay for all the reactions after an order sent by the synchronizer in EMN to simultaneous firing of thousands electrocytes in each electroplate, is less than 0.5 ms.

3. Discussion

The principle of generating EM pulses is the same in all biosystems, which is ruled by Maxwell Equations, that a transient transmembrane ionic current ($Na^+$, $K^+$ or $Ca^{2+}$) should generate a pulse [19, 20]. The EM pulses transmit in dielectric materials in a way like solitons, which keeps their shape, amplitude and velocity almost unchanged for a long distance [5, 16, 17].

From above analyses, we may figure out a common feature in electrical communication in various biosystems: A softmaterial waveguide network universally exists, and this network serves efficiently as the propagation path for EM pulse in biosystems. Lipid membranes are the main frames for the network, and sieve tubes and plasmodesma in plants and myelin sheath in myelinated axons, also play the same role. Together with ionic solutions inside and outside of cells, they become sandwich-structured waveguides. The cases of paramecia, flytraps and electric eels show that this network could work with an extremely high efficiency.

The fact that electric eels are able to make high-voltage, high-current hunting discharge within 0.1-0.5 ms is a unique case to verify different theories or hypotheses for electrical communication in biosystems. Indeed Hodgkin and Huxley model is not suitable to explain the internode transmission of action potentials in myelinated axons, where transient transmembrane ion flows at one node can trigger voltage gated ion channels in neighboring node 1-2 mm away. This is because of the screen effect, that rearrangement of the locations of ions at one node cannot create transversal an electric field along the axon for initiating an electrical (ionic) current. Therefore, Hodgkin - Huxley model fails to explain the high propagation speed (100-150 m/s) of electrical signals in myelinated axons, neither to explain the extreme high signal speed observed in the EMN of electric eels. In electromechanical solitons theory, the propagation velocity of signals in dipalmitoyl phosphatidylcholine unilamellar vesicles were calculated to be in the order of 100 m/s. This is also much lower than what needs in the case of simultaneous discharge activities in electric eels. It addition, in soft lipid membranes a mechanical wave is supposed to attenuate much faster over distance than an EM pulse does. As discussed previously, an EM pulse travels in dielectric membranes in a speed of the order of $10^8$ m/s. So in the model of EM pulse and softmaterial waveguide network, delay in propagation of the EM pulses is nearly 100% caused by the time cost in transmembrane ion flows over opened protein channels and number of relays over the propagation path [20]. It seems that over the long history of evolution, electric eels have developed unique membrane network structures for their synchronizer in EMN and discharging electroplate in electric organ, as well as for descending fibers connecting these parts, so as to realize a minimum transmission delay of the EM pulses.

## 4. Conclusion

By analyzing the "simultaneous phenomena" observed in typical cases of unicellular organisms, plants and animals, where a big amount reactors in a biosystem almost simultaneously response to a single input signal and complete reactions within milliseconds, we have come to the conclusion that biosystems have developed softmaterial waveguide networks and utilized soliton-like EM pulses in these extremely fast and synchronized activities, no matter with or without the existence of neurons and axons. We have further addressed that, the membrane-based waveguide networks and ion-channel induced EM pulses universally exist in biosystems at many levels, from the whole body level such as an electric eel, to sub-cell level such as cell organelles. For generation of the EM pulses, it needs a local gradient in transmembrane ion concentration and ion channels. Increase in thickness of the membrane enhances the transmission efficiency for the EM pulses propagating over a long distance, and it needs more ion channels working as relays on the way to reproduce the signals.

We have probably revealed the nature of "wireless communication" in biosystem. At difference levels, millions and billions of molecules, cell organelles and cells find their way to communicate and work coordinately and even synchronously, to complete complicated bio-activities and performances. Under this point of view, neural systems are specially developed organs for electrical communication at higher efficiency than that in normal membrane network among cells and cell organelles. This work may offer an alternative view and paradigm for analyzing the synchronized working mechanism of ten billion neurons in a human brain.

**Acknowledgments**

We thank Prof. M. Z. Li, Prof. H. B. Han, Dr. J. Tang and R. J. Dai for valuable discussions.